\documentclass[aps,prb,prbbib,twocolumn,epsf]{revtex4}
        \usepackage{graphicx}
\usepackage{bm}
\usepackage{amsmath}
\usepackage{setspace}
\begin{document}
\title{Current induced torques between ferromagnets and compensated antiferromagnets: symmetry and phase coherence effects}
\author{Karthik Prakya$^{1}$, Adrian Popescu$^{2,3}$, and Paul M. Haney$^{2}$}
\affiliation{
1.  The MITRE Corporation, Bedford, MA 01730  \\
2.  Center for Nanoscale Science and Technology, National Institute of Standards and Technology, Gaithersburg, MD 20899 \\
3.  Maryland NanoCenter, University of Maryland, College Park, MD 20742, USA
}

\begin{abstract}
It is shown that the current-induced torques between a ferromagnetic layer and an antiferromagnetic layer with a compensated interface vanish when the ferromagnet is aligned with an  axis of spin-rotation symmetry of the antiferromagnet.  For properly chosen geometries this implies that the current induced torque can stabilize the out-of-plane (or hard axis) orientation of the ferromagnetic layer.  This current-induced torque relies on phase coherent transport, and we calculate the robustness of this torque to phase breaking scattering.  From this it is shown that the torque is not linearly dependent on applied current, but has an absolute maximum.
\end{abstract}

 \maketitle

\section{Introduction}

Current-induced torques result from the interaction between conduction electron spins and the magnetization of a sample when current flows through it.  This torque is generally present when the magnetization is spatially nonuniform, and has been extensively studied in the context of magnetic domain walls and spin valve structures.  Since its theoretical prediction\cite{slonc,berger}, extensive studies have led to a theoretical framework of current-induced torque in ferromagnets that describes experimental results with quantitative success.\cite{stiles:jmmm}  It has been proposed that current-induced torques also exist in antiferromagnetic systems.\cite{nunez,haney:jmmm}  Previous theoretical studies considered systems composed entirely of antiferromagnetic layers \cite{nunez,duine:prb,haney:prb} as well as experimental \cite{wei} and theoretical\cite{haney:prl,loktev,loktev2} systems with both ferromagnetic and antiferromagnetic layers.  Theoretical work has also focused on antiferromagnet textures.\cite{brataas1,brataas2,duine:2011,niu} Experiments have demonstrated current-induced torque in materials with other types of complex magnetic ordering, such as skyrmion lattices.  Recent theory\cite{shick} and experiment\cite{jungwirth} have shown that antiferromagnets exhibit anisotropic magnetoresistance, demonstrating a coupling between magnetic order and charge transport these materials.

Antiferromagnets exhibit an array of magnetic ordering, such as spin density waves that are commensurate or incommensurate with the lattice, and configurations with multiple spin density waves.  As shown in Ref. \onlinecite{haney:prl}, the symmetry properties of the antiferromagnetic layer can lead to torques in multilayers with qualitatively different properties than conventional spin valves.  In particular, a collinear compensated antiferromagnetic layer interface (with each spin in the $\pm {\hat z}$ direction, which we call a 1Q spin structure) leads to a torque which vanishes when the ferromagnet is perpendicular to the $\hat z$ direction.  This torque can stabilize the hard-axis orientation of the ferromagnet in systems where the antiferromagnet is pinned.  Here we treat similar systems (see Fig. 1a), and compute the current-induced torque on the ferromagnetic layer.  (Previous works have investigated the current-induced torque on the antiferromagnetic layer in such systems.\cite{loktev,loktev2})

In this work we consider a system where the antiferromagnetic layer has a 3Q spin structure (see Fig. 1b).  This is qualitatively different than the previously studied 1Q antiferromagnet because the 3Q structure has only a {\it single} axis of spin rotational symmetry (3-fold in this case), whereas for the 1Q antiferromagnet {\it all} directions perpendicular to the $\hat z$ direction are axes of 2-fold spin rotational symmetry.  We show that an important consequence of the reduced symmetry of the 3Q antiferromagnet is that the current-induced torque stabilizes the out-of-plane magnetic orientation only when the magnetization is initialized nearby this orientation (in contrast, the 1Q antiferromagnet drives {\it any} initial orientation out-of-plane).  In this work we additionally determine the effects of phase breaking scattering: The current-induced torque relies on phase coherence, and quantifying the robustness with respect to scattering is important to gauge the feasibility of observing these effects in real systems.

Our results are easily generalized to multilayers composed of a free ferromagnet layer, and a fixed magnetic layer whose spin configuration has an axis of $n$-fold rotational symmetry.  The key property of the torque is that: if the ferromagnetic layer is aligned with an axis of spin-rotational symmetry of the fixed layer, then the current-induced torque (in fact, all torques) must vanish.  This is seen by recognizing that, by assumption, the system is invariant with respect to spin rotations about the ferromagnet orientation by some angle $\phi_n$, and any nonzero torque (which must be perpendicular to the ferromagnet orientation) does not respect this symmetry.  For conventional spin valves, this statement implies the well known fact that the current-induced torque vanish when the ferromagnet layers are aligned or anti-aligned.  Identifying the points where the current-induced torque vanishes is important because the torque may drive the magnetization to these fixed points.  For properly designed antiferromagnet-ferromagnet multilayers this property of the torque can stabilize the out-of-plane magnetic orientation.\cite{haney:prl}  This is because this orientation, being a maximum of the magnetic free energy, represents a fixed point for the conventional micromagnetic torques.  In the absence of current-induced torques, this out-of-plane configuration is an unstable fixed point; however if the current-induced torque drives the ferromagnet to this orientation and exceeds the damping torque, it can stabilize this configuration, as shown by micromagnetic simulations in Ref. \onlinecite{haney:prl}.

\section{Method}\label{sec:method}

To calculate the current-induced torques, we use the nonequilibrium Green's function technique within a tight binding representation.  This is a well established approach to calculating the transport properties of magnetic thin films.  We highlight the most important details here.  The system is taken to consist of two semi-infinite electrodes, with a scattering region placed between them.  There is a difference $V_{\rm app}$ in the electrochemical potential of the two electrodes.  The central quantity is the density matrix $\rho$:

\begin{eqnarray}
\rho &=& \frac{i}{2 \pi} \int_{-\infty}^{E_{\rm F}-V_{\rm app/2}} \left[G^r\left(E\right)-G^a\left(E\right)\right] dE +  \nonumber\\
&&~~~\int_{E_{\rm F}-V_{\rm app}/2}^{E_{\rm F}+V_{\rm app/2}} G^r\left(E\right) \Gamma_L \left(E\right) G^a\left(E\right) dE. \label{eq:rho}
\end{eqnarray}
where $G^{r,a}\left(E\right)=\left(E-H_C-\Sigma^{r,a}_L\left(E\right)-\Sigma^{r,a}_R\left(E\right)\right)^{-1}$.  $H_C$ is the scattering region Hamiltonian, and $\Sigma^r_{L}$ is the self energy which describes the electronic coupling between the scattering region and the semi-infinite left lead; it is given by $\Sigma^r_{L} = V_{C,L}^\dagger g_{0,L}^r\left(E\right)V_{C,L}$, where $V_{C,L}$ is the coupling matrix element between the left lead and central region, and $g_{0,L}$ is the surface Green's function of the isolated semi-infinite left lead.  The same form of self energy holds for the right lead.

As noted in previous works,\cite{duine:prb} phase coherence plays a central role in a number of the antiferromagnetic systems studied so far.  To explore the robustness of the torques in this system, we include an additional self energy $\Sigma_S$ in Green's function which describes elastic, phase breaking scattering.  Its form is:
\begin{eqnarray}
\Sigma_S\left(E\right) = i D \left(G^r\left(E\right) - G^a\left(E\right)\right)
\end{eqnarray}
where $D$ parameterizes the scattering.  A discussion of the parameter $D$ in terms of real material properties and temperature is given in Sec. (\ref{sec:results}).

We assume the spin-orbit coupling is negligible, so that the current-induced torque on the ferromagnet layer is simply given by the transverse component of incoming spin current flux.  For our geometry, the net spin current has real space velocity in the $\hat y$ direction.  The spin current operator $\vec{J}\left(y\right)$ is then:

\begin{eqnarray}
\hat{\vec J}\left(y\right) =  \sum_{\substack{j\in R\left(y\right)\\ k\in L\left(y\right)\\ s,s'}} i \left[ c^\dagger_{j,s} \vec {\sigma}_{s,s'} c_{k,s'} t_{j,k} - {\rm h.c.} \right],
\end{eqnarray}
where $R\left(y\right)$ are the set of sites with coordinate $y'$ greater than $y$, and $L\left(y\right)$ are the set of sites with coordinate $y'$ less than $y$.  $\vec\sigma$ sigma is the vector of Pauli matrices, and we take the hopping $t_{j,k}$ between all sites $j$ and $k$ to be spin independent.  We present results in terms of torque per current (units of ${\mu_B}/e$), which represents the spin torque efficiency.  The absolute value of this quantity determines the critical current needed to drive magnetic dynamics.

As discussed in Refs. \onlinecite{haney2007} and \onlinecite{nikolic}, it is sometimes necessary to compute the entire energy integral (both terms in Eq. \ref{eq:rho}) in order to find the current-induced torques.  This is particularly the case when the torques in question are present in equilibrium (which is itself dependent on the symmetries of the system, as discussed in Ref. \onlinecite{haney2007}).  We checked explicitly that the current-induced torque in question for these systems are dominated by the nonequilibrium contribution to the density matrix (the second term of Eq. \ref{eq:rho}), and present only this contribution in the results (the remaining ``energy integral" contribution is several orders of magnitude smaller in all the cases we checked).  We take the Fermi energy to be $E_F=3.75~t$, and use a dense $k$-point mesh to converge the transport integrals, up to $1000^2$ k-points for a unit cell having 4 atoms per layer.

\begin{figure}
\includegraphics[width=1\columnwidth]{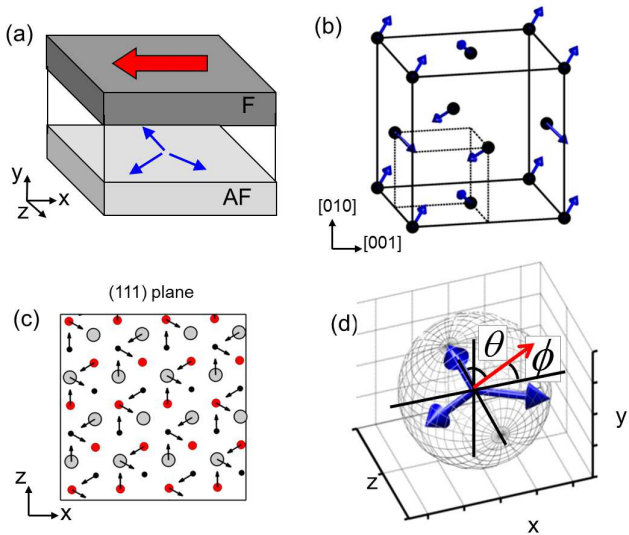}
\caption{(a) Overall system geometry (b) The crystal and spin structure for the 3Q state.  The spins at the corners of the interior box all point inward. (c) The spin on the [111] interface of the lattice from (b).  The small black, medium red, and large gray dots represent atoms in different layers ({\it i.e} with different y-values).  The spin of dots without an arrow is completely in the $\hat{y}$ direction, while other spins are partially canted in the ${\hat y}$ direction. (d) Spherical coordinate system used to describe the torques on the ferromagnet.  The blue (dark) spins in the $x-z$ plane represent the 3-fold symmetric spins of the antiferromagnetic layer, and the skinnier red arrow represents the orientation of the ferromagnet layer. }\label{fig:sys}
\end{figure}

A schematic of the overall system is shown in Fig. \ref{fig:sys}a.  It consists of semi-infinite ferromagnetic and antiferromagnetic layers, separated by a nonmagnetic spacer which is 3 atomic layers thick. The layers are fcc, with interfaces in the [111] direction.  We use two different spin structures for the antiferromagnet.  One is a 3Q spin structure, depicted in Fig. \ref{fig:sys}b.  The spin structure in the $(111)$ planes is shown in Fig. \ref{fig:sys}c, which shows the 3-fold symmetry of the spin in the $x-z$ plane.  Each spin also has a component along the $y$ axis (into or out of the page).  Atoms with no arrow in the figure have a spin fully aligned in the $+\hat{y}$ direction, while other atoms' spins are partially canted in the $-\hat{y}$ direction, so that the net bulk spin vanishes.  Common antiferromagnetic materials such as FeMn are predicted to have a 3Q ground state,\cite{schulthess,footnote2,stocks}  consistent with measurements\cite{kawarazaki,kennedy}, although there is not complete consensus between all the experimental data.  To further explore the consequences of the antiferromagnet symmetry, we also consider a system where the $y$ component of the spins are set to 0.   This artificial system retains the 3-fold symmetry in the $x-z$ plane, but is also symmetric under $s_y\leftrightarrow-s_y$.  We refer to this as the ``no-canting" antiferromagnet.  We emphasize that our primary results generalize to any antiferromagnet for which there is an axis of $n$-fold spin rotational symmetry, as explained in the introduction.

We present the angular variation of the torque on the ferromagnet layer in terms of spherical coordinates, as shown in Fig. 1d.  The $\hat y$ direction is the hard axis of the F, which is taken to coincide with the axis of 3-fold symmetry of the antiferromagnet.  As explained in the introduction, this alignment of hard axis and the antiferromagnet axis of spin rotational symmetry is crucial for the out-of-plane orientation to be stabilized by the current-induced torque. The $\hat z$ direction is along one of the spins of the antiferromagnetic layer.  We utilize similar schematics as Fig. \ref{fig:sys}d in the next section to show the relative orientation of the ferromagnet layer with the spins of the antiferromagnet.

\section{Results}\label{sec:results}

\begin{figure}
\includegraphics[width=1\columnwidth]{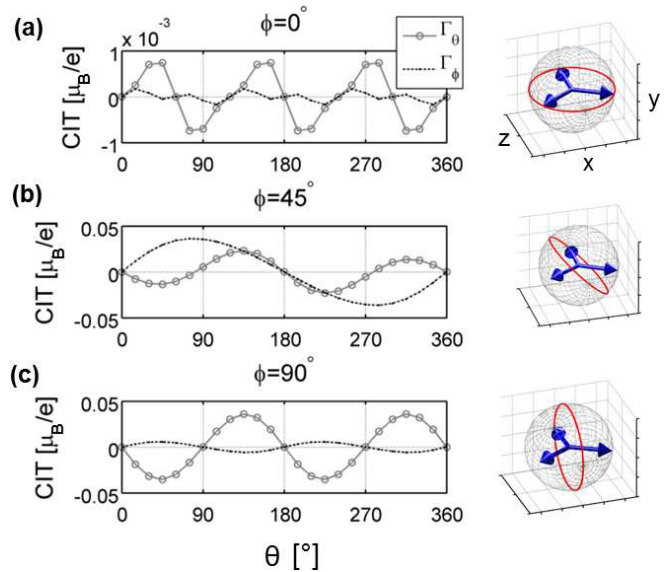}
\caption{The angular dependence of the current-induced torque (CIT) on the ferromagnet for the system with no antiferromagnetic canting in the $y$-direction (the ``no canting" system).  The black dashed line is the torque in the $\hat{\phi}$ direction, and the gray line with markers is the torque in the $\hat \theta$ direction. (a) shows the torque as when the ferromagnet is coplanar with the antiferromagnet spins, which shows a $\sin(3\theta)$ dependence.  (b) shows an intermediate angle, and (c) shows the torque as the ferromagnet orientation is normal to the plane of the antiferromagnet spins.  In this case, the torque varies as $\sin\left(2\theta\right)$.  The diagrams to the right of the plots show the direction of antiferromagnet spins in the x-z plane, and with a circle representing the angles of the ferromagnet layer in the plot.}\label{fig:nocant}
\end{figure}

The current-induced torque on the ferromagnetic layer for a no-canting antiferromagnetic system is shown in Fig. \ref{fig:nocant}.  Unlike the current-induced torque in a conventional spin valve, whose magnitude has a simple $\sin(\theta)$ dependence, we find a more complex angular dependence for the torque.  We first fix $\phi=0^\circ$ and vary the ferromagnet orientation from $\theta=0$ to $360^\circ$.  These orientations are in the easy plane.  The torques conform to the 3-fold symmetry, varying approximately as $\sin\left(3\theta\right)$, as shown in Fig. \ref{fig:nocant}a.  For fixed $\phi=90^\circ$, sweeping the polar angle $\theta$ takes the magnetization out of the easy plane, through the hard axis direction.  The torques in this case are shown in Fig. \ref{fig:nocant}c. The torques vary as $\sin\left(2\theta\right)$, again as required by symmetry.  For fixed $\phi=45^\circ$, varying $\theta$ takes the ferromagnet on an ``off-axis" orbit, and the torque exhibits more complex angular dependence.

\begin{figure}
\includegraphics[width=1\columnwidth]{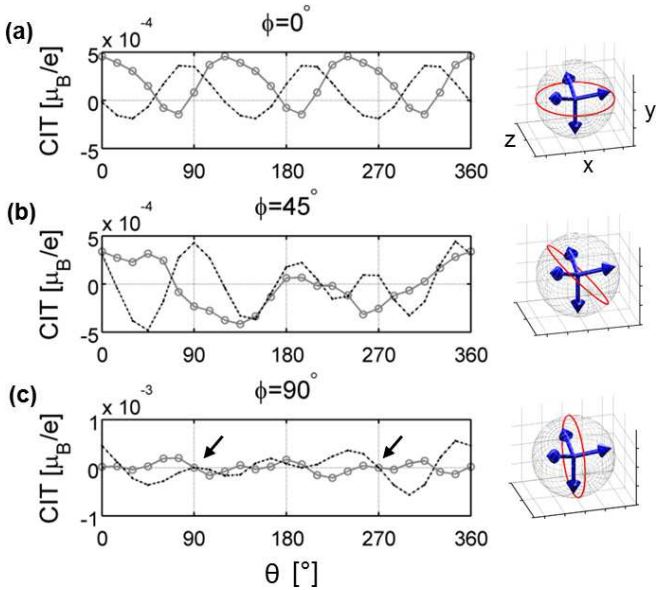}
\caption{The angular dependence of the torque on the ferromagnet for the system with no 3Q spin ordering of the antiferromagnet.  The black dashed line is the torque in the $\hat{\phi}$ direction, and the gray line with markers is the torque in the $\hat \theta$ direction. (a) shows that the torque again varies as $\sin(3\theta)$ when the ferromagnet layer is confined to the $x-z$ plane (easy plane).  (b) shows complex angular dependence for the ferromagnet layer oriented along an axis of low symmetry. (c) shows that the torque vanishes when the ferromagnet is aligned to the axis of 3-fold symmetry of the antiferromagnet (arrows indicate these points).}\label{fig:cant}
\end{figure}

 Figure 3 shows similar results for the 3Q spin structure for the same set of magnetic orientations.  The reduction in symmetry due to the inequivalence of $s_y$ and $-s_y$ leads to more complex behavior of the torque.  For $\phi=0^{\circ}$, we note the invariance of the torque under $\theta \rightarrow \theta+120^{\circ}$.  Key data points are shown in Fig. \ref{fig:cant}c by the black arrows.  As argued in the introduction, when the ferromagnet layer is aligned to the axis of 3-fold symmetry, the current-induced torque vanishes.

\begin{figure}
\includegraphics[width=1\columnwidth]{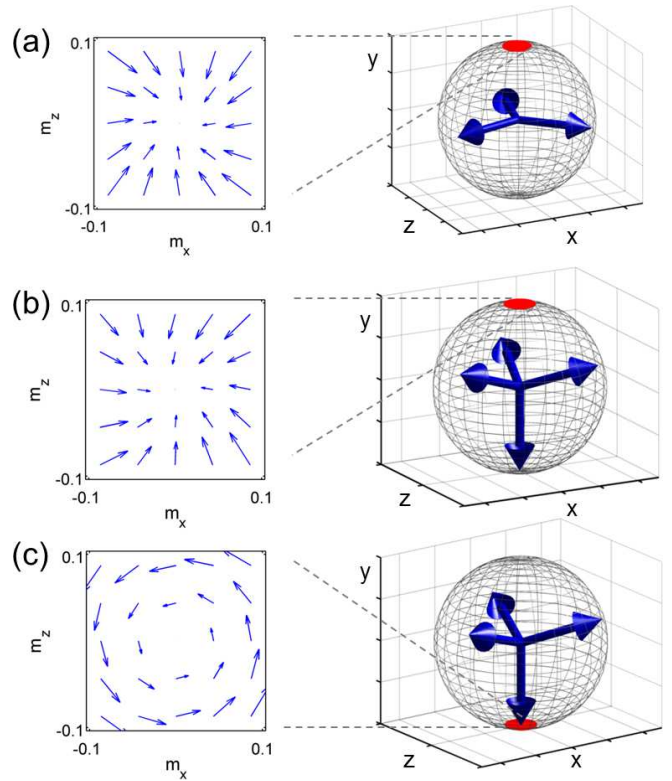}
\caption{A zoom-in view of the torques on the ferromagnet layer near the fixed point of the current-induced torque. (a) shows the result for the ``no canting" system, where the $\pm y$ fixed points are equivalent.  The red dot on the sphere on the right represents the magnetic orientation shown in the left panel.  The dark blue arrows represent the orientation of the antiferromagnet spins. (b) shows the result for the 3Q system.  The torques indicate that the $+{\hat y}$ orientation is a stable fixed point.  (c) shows that the $-{\hat y}$ orientation is an elliptic fixed point.  (The three blue (dark) arrows of (a) have no $\hat y$ component, while for (b) and (c), the three similar blue (dark) arrows are canted, acquiring a small positive $\hat y$ component.)}\label{fig:fp}
\end{figure}

To gain a fuller view of the current-induced torque near the out-of-plane fixed point, we show the torque in the vicinity of these points in Fig. \ref{fig:fp}.  For the no-canting system, the $+\hat{y}$ and $-\hat{y}$ fixed point are equivalent.  For electrons flowing from the antiferromagnet to the ferromagnet, these are stable fixed points.  For the 3Q antiferromagnet, on the other hand, the $+\hat{y}$ and $-\hat{y}$ fixed points are inequivalent.  In this case, we find the $+\hat{y}$ is a stable attractor, while the $-\hat{y}$ is an elliptic fixed point.  The nature of the fixed point (stable, unstable, elliptic, etc.) is parameter dependent, making it difficult to make general statements about the prevalence of different fixed points.

For antiferromagnetic systems it is also important to distinguish between stable fixed points to which any initial magnetization vector is driven (global attractors), and those fixed points for which only an initial magnetization vector nearby is driven (local attractors).  Inspection of Fig. \ref{fig:nocant}a shows that, if the magnetization is in the $x-z$ plane, the torque driving it to the out-of-plane direction is quite weak (in this case, the relevant torque is in the $\hat{\phi}$ direction).  On the other hand, if the magnetization is near the $z-y$ plane (Fig. \ref{fig:nocant}c), the torque driving it to the out-of-plane orientation ($\Gamma_\theta$) is much stronger.  Rather than characterizing the flow of the current-induced torque field for any particular system in detail (which is highly parameter dependent), we simply emphasize that an experiment is more likely to observe these torques if the magnetization is initially in the out-of-plane before the current is applied.  Application of a current can stabilize this configuration, so that subsequent removal of the applied field does not result in the magnetization returning to the easy plane.

\begin{figure}
\includegraphics[width=1\columnwidth]{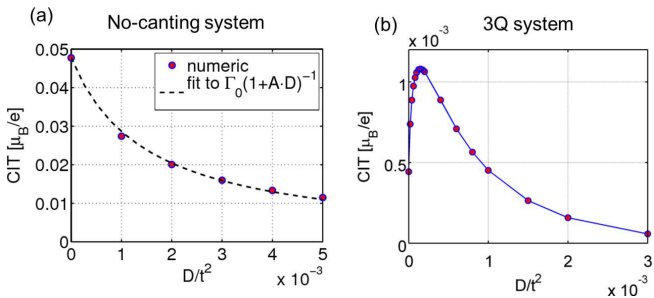}
\caption{(a) The magnitude of current-induced torque near the fixed point of the ``no-canting" system as a function of the elastic scattering parameter ${D/t^2}$.  The parameters used in the curve fit are: $\Gamma_0=0.0478~\left(\mu_B/e\right),~A=670~t^{-2}$.  (b)  The same plot for the 3Q system. The fit applies only to (a).}\label{fig:scattering}
\end{figure}

In contrast to the current-induced torque in noncollinear ferromagnets, the current-induced torque in many antiferromagnet systems rely on phase coherence.\cite{duine:prb}  This is because the eigenstates of the bulk antiferromagnet are degenerate Kramer's doublets with opposite spins.  A distribution of these eigenstates carries no net spin current.  However, spin-dependent reflection at the ferromagnet interface leads to a superposition of these degenerate states, which results in a nonzero spin polarization of the current in the antiferromagnet.  The component of this spin current perpendicular to the ferromagnet is responsible for the torque on the F, and vanishes as the coherence between the states is destroyed.  The requirement of ballistic (or quasi-ballistic) transport imposes more stringent requirements on the existence of current-induced torques in antiferromagnets than in ferromagnets.  Materials should be nearly single crystal, and scattering (from e.g. phonons) should be minimized.  In order to estimate the acceptable limits of electron-phonon scattering, we add an elastic scattering channel to the Green's function self-energy as described in Sec. \ref{sec:method}.  Fig. \ref{fig:scattering} shows how increased scattering decreases current-induced torque near the out-of-plane fixed point of the no-canting system.  Here the scattering parameter $D$ is normalized by the square of the hopping matrix element $t$.

To place the result of Fig. \ref{fig:scattering} in context, we write $D$ in terms of material properties.  For simplicity, we focus on just one phase breaking process: elastic acoustic phonon scattering.  Our aim is to explicitly show that the current-induced torque, as a function of the applied current, has a maximum absolute value.  Depending on materials properties and temperature, other scattering processes may be more important. In any event, for acoustic phonon scattering, $D$ takes the form:\cite{lundstrom}
\begin{eqnarray}
D = \frac{E_a^2 k_{\rm B}T}{\rho v^2 a^3} \equiv D_0 T \label{eq:D}
\end{eqnarray}
where $E_a$ is the elastic deformation potential, $\rho$ is the material density, $v$ is the speed of sound, $a$ is the lattice spacing, and $T$ is the temperature.  The linear $T$ dependence reflects the increased thermal population of phonons with increasing temperature.  Other scattering process ({\it e.g.} electron-electron scattering, inelastic phonon scattering) depend on $T$ differently; generally $D \propto T^p$ where $p$ varies from 0.5 to 3 (see Ref. \onlinecite{mohanty} and references within).

Joule heating may increase the importance of thermal effects:  for current density $J$ flowing through a material with resistivity $\Omega$, thermal conductivity $\kappa$, and length $L$ along the current direction (in this case, the $\hat y$-direction),  the spatially averaged temperature increases by a factor on the order of $J^2L^2 \Omega / \kappa$.  To stabilize the out-of-plane magnetic orientation requires a current density of $\alpha \gamma M_s t_{\rm F}/2g$\cite{footnote1}, where $g$ is the current-induced torque per current, $\alpha$ is the damping, $\gamma$ is the gyromagnetic ratio, $M_s$ is the saturation magnetization of the ferromagnet layer, and $t_{\rm F}$ is the thickness of the ferromagnet layer.  For the no-canting system, the current-induced torque per current is $g=0.05~\mu_{\rm B}/e$.  According to this estimate and typical material parameters, this leads to a critical current density on the order of $10^{12}~{\rm A/m^2}$.  Taking $\rho=10^{-7}~{\rm\Omega\cdot m},~\kappa=50~{\rm W/\left(m\cdot K\right)},L=50~{\rm nm}$ leads to only a modest increase in temperature, less than $10~{\rm K}$.  The other parameters of Eq. \ref{eq:D} for metals are typically $E_a=10~{\rm eV},~\rho=10^4~{\rm kg/m^3},~v=5000~{\rm m/s},~a=0.35~{\rm nm}$.  In total, we find a $D$ parameter on the order of $10^{-5}~{\rm eV}^2$ to $10^{-4}~{\rm eV}^2$.  In light of Fig. \ref{fig:scattering}, this implies that elastic phonon scattering does not immediately destroy the current-induced torque for the no-canting system.  On the other hand, the much weaker current-induced torque per current of the 3Q system ($g=4\times10^{-4}~\mu_{\rm B}/e$) requires a 100-fold increase in the current to stabilize the out-of-plane orientation, a current density which exceeds the maximum these systems can accommodate.

We've observed that the current-induced torque decays as $1/D$ for the no-canting system.  This is not universal behavior.  Indeed, the current-induced torque in the 3Q system is nonmonotonic with scattering parameter $D$.\cite{footnote3}  Despite its non-universality, we find it instructive to assume such a dependence in order to derive closed form expressions for the maximum current-induced torque as a function of applied current density.  Recalling that $D$ is proportional to $T$, we find the {\it absolute} current-induced torque $\Gamma_{\rm abs}$ (units of torque) varies with current as:
\begin{eqnarray}
\Gamma_{\rm abs}\left(J\right) = \frac{\Gamma_0 J}{1+A D_0\left( T_0 + B J^2\right)},\label{eq:tj}
\end{eqnarray}
where $\Gamma_0$ is the current-induced torque in the absence of scattering (recall $\Gamma_0$ has units of torque per current), $T_0$ is the sample temperature in the absence of current, $B=L^2\rho/\kappa$ describes the system's susceptibility to current-induced heating, and $D_0$ is defined in Eq. \ref{eq:D}.\cite{footnote2}  The absolute current-induced torque has a maximum - for current densities that are too large, the magnitude of the current-induced torque decreases due to increased scattering from Joule heating.  The maximum absolute current-induced torque is given by:
\begin{eqnarray}
\Gamma_{\rm abs}^{\rm max} = \frac{\Gamma_0}{3L} \sqrt{\frac{2\kappa}{\rho D_0 A\left(1+D_0 T_0 A\right)}},\label{eq:max}
\end{eqnarray}
 The parameters $\Gamma_0$ and $A$ are entirely system specific, and related to the spin dependent transport properties of a system, and their robustness with respect to scattering.  If the above maximum torque exceeds the damping torque $\alpha \gamma M_s t_{\rm F}/2g$, then the out-of-plane configuration can be stabilized by the current-induced torque.  Intuitively, it's advantageous to use a low $M_s$ material in order to reduce the critical current, and a thin multilayer to reduce heating.  For scattering processes with different functional dependence on $T$, a similar line of reasoning applies, although the specific form of the maximum absolute current-induced torque will differ.  It's straightforward to show that a $T^p$ dependence of $D$ results in a maximum current-induced torque expression similar to Eq. \ref{eq:max}, where the expression inside the square root is taken to the power $1/2p$.

\section{Conclusion}
This work demonstrates the role of symmetry and phase coherence effects in the current-induced torque present between ferromagnet and antiferromagnetic layers with a compensated interface.  Basic symmetry arguments identify the fixed points of the current-induced torque.  We demonstrate that for an antiferromagnetic layer with a 3Q spin structure, the current-induced torque has a complex angular dependence, and the fixed points for the current-induced torque are generally only local attractors.  This is important because experiments designed to drive the ferromagnet to these fixed points must initialize the ferromagnet sufficiently nearby.  We also show via explicit calculations the primary role played by phase coherence for these torques, and show an inverse relationship between the magnitude of the current-induced torque and the phase breaking scattering parameter.  In the antiferromagnetic system with planar spins (the no-canted system), we find the current-induced torque to be sufficiently robust to scattering to stabilize the out-of-plane magnetic orientation, while for the 3Q ordered antiferromagnet, the current-induced torque is too weak to stabilize this orientation.  We expect that the robustness of this torque to scattering should be system specific, determined by which scattering processes are dominant and the system electronic structure.

\section{Acknowledgements}
A.P. acknowledges support under the Cooperative Research
Agreement between the University of Maryland and the
National Institute of Standards and Technology Center for
Nanoscale Science and Technology, Award 70NANB10H193,
through the University of Maryland.


\begin{thebibliography}{00}
\bibitem{footnote2} We note that the true ground state may differ slightly from the 3Q configuration in the case of FeMn, as shown in Ref. \onlinecite{stocks}.  We expect that fluctuations or deviations from ordered configurations will decrease the effectiveness of symmetry-based torques.
\bibitem{footnote1} This expression for the current density required to stabilize the out-of-plane orientation assumes that there is no applied magnetic field, or other sources magnetic anisotropy.  In this case, the damping torque from the hard-axis anisotropy for a magnetization with small tilt angle $\beta$ away from the hard-axis is $\gamma\alpha M_s \beta$, while the current-induced torque is $2gJ\beta/t_{\rm F}$.  Equating these two leads to the form of current density given in the text.
\bibitem{footnote2} We assume that the material paramters in Eq. \ref{eq:D} are weakly temperature dependent in this analysis.
\end{thebibliography}
\end{document}